\documentclass[a4paper, 10pt]{article}
\usepackage{mathrsfs}
\usepackage{bbm}
\usepackage{amssymb}

\usepackage[dvipdfmx]{graphicx}
\usepackage{authblk}

\usepackage[top=30truemm,bottom=30truemm,left=30truemm,right=30truemm]{geometry}

\title{\large{\textbf{Simple Stochastic Order-Book Model of Swarm Behavior \\ in Continuous Double Auction}}}

\author[1,2]{\normalsize{Shingo Ichiki}}
\author[3,4]{\normalsize{Katsuhiro Nishinari}}

\affil[1]{\small{Department of Advanced Interdisciplinary Studies, Graduate School of Engineering, The University of Tokyo, 4-6-1, Komaba, Meguro-ku, Tokyo 153-8904, Japan}}
\affil[2]{\small{Tokyo Stock Exchange, Inc., 2-1, Nihombashi-kabuto-cho, Chuo-ku, Tokyo 103-8220, Japan}}
\affil[3]{\small{Research Center for Advanced Science and Technology, The University of Tokyo, 4-6-1, Komaba, Meguro-ku, Tokyo 153-8904, Japan}}
\affil[4]{\small{Department of Aeronautics and Astronautics, Graduate School of Engineering, The University of Tokyo, 7-3-1, Hongo, Bunkyo-ku, Tokyo 113-8656, Japan}}

\date{\small{November 8, 2014}}

\makeatletter
\renewcommand{\section}{%
  \@startsection{section}%
   {1}%
   {\z@}%
   {-3.5ex \@plus -1ex \@minus -.2ex}%
   {2.3ex \@plus.2ex}%
   {\normalfont\normalsize\bfseries}%
}%
\makeatother

\makeatletter
\long\def\@makecaption#1#2{%
  \small
  \vskip\abovecaptionskip
  \sbox\@tempboxa{#1: #2}%
  \ifdim \wd\@tempboxa >\hsize
    #1: #2\par
  \else
    \global \@minipagefalse
    \hb@xt@\hsize{\hfil\box\@tempboxa\hfil}%
  \fi
  \vskip\belowcaptionskip}
\makeatother

\begin{document}
\maketitle

\begin{abstract}

In this study, we present a simple stochastic order-book model for investors' swarm behaviors seen in the continuous double auction mechanism, which is employed by major global exchanges. Our study shows a characteristic called ``fat tail" is seen in the data obtained from our model that incorporates the investors' swarm behaviors. Our model captures two swarm behaviors: one is investors' behavior to follow a trend in the historical price movement, and another is investors' behavior to send orders that contradict a trend in the historical price movement. In order to capture the features of influence by the swarm behaviors, from price data derived from our simulations using these models, we analyzed the price movement range, that is, how much the price is moved when it is continuously moved in a single direction. Depending on the type of swarm behavior, we saw a difference in the cumulative frequency distribution of this price movement range. In particular, for the model of investors who followed a trend in the historical price movement, we saw the power law in the tail of the cumulative frequency distribution of this price movement range. In addition, we analyzed the shape of the tail of the cumulative frequency distribution. The result demonstrated that one of the reasons the trend following of price occurs is that orders temporarily swarm on the order book in accordance with past price trends.
\end{abstract}

\section{Introduction}
Numerous stocks and derivatives are traded on exchanges around the world. For the traders, it is crucial to estimate the profitability and downside risk of financial products they trade in. In the past, miscalculation of price movements has caused many global financial crises. Long-Term Capital Management, a hedge fund management firm, is one of the companies that has caused such crises when it suddenly incurred enormous loss and recapitalized in 1998. One of the reasons for its recapitalization was because the prices of financial products substantially changed unexpectedly. R. N. Mantegna and H. E. Stanely have pointed out, prices actually do not ideally move following a normal distribution \cite{rev17}. In reality, large-scale price movement more frequently occurs than that assumed by a normal distribution. Various financial markets have experienced large-scale price movement that does not fit a normal distribution, and studies of market crashes are flourishing \cite{rev4}.

In this study, we studied large-scale price movements in the exchange market caused by investors' collective behaviors. Collective behaviors can be seen in various settings including economic and biological as well as other social setting. In recent years, this collective behavior has been termed ``swarm intelligence." These behaviors have been extensively studied as phenomena that show a high degree of motion as a group through simple local interactions between individuals \cite{rev5}. As pioneering research on collective behavior in financial markets, T. Lux and M. Marchesi explored models to reproduce phenomena, such as fat tails and the temporal independence of volatility, with regard to the interaction of strategic investor groups \cite{rev19}. The present study focuses on the how local investors affect each other and produce large-scale price fluctuations as a group. As such, we herein denote investors' collective behavior as ``swarm behaviors." We think one of the factors of large-scale price movement is connected with certain swarm behaviors of investors. For example, once the price begins to continuously decline, investors fear downside risk, and they follow the trend by lowering the prices of their orders. This is one of the mechanisms by which price decline causes further declination, leading to an extreme price fall. Y. Hashimoto et al. used actual foreign exchange market data and indicated that the market price is dependent on past price trends \cite{rev21}. Thus, we think it is meaningful to focus on the phenomena brought about by investors' swarm behaviors. In particular, in the present study, we investigate two swarm behaviors: one that follows a trend in the historical price movement that is seen in general financial market and another that contradicts a trend in the historical price movement. We present a simple stochastic order-book model that captures these swarm behaviors of investors. From the price data derived from simulations using these models, we capture the behaviors of investors. We analyze the cumulative distribution of the price movement range when the price continuously move in a single direction.

\section{Model}
In this section, we describe the structure of our proposed model. Our model is applied to trading on exchanges; therefore, we begin by explaining the trading structure used by major exchanges. Major global financial exchanges use the continuous double auction mechanism. The key idea of this mechanism is the use of an electronic board called an order book where each trader writes bids and asks on and orders are matched on. In addition, two main order types are used by major exchanges. One is the limit order used to indicate a price upon which traders wish to have their orders executed, and another is the market order that is used without an indication of price. When sending an ask limit order, if there is a bid order on the order book that has the same or higher price than a trader's ask price, then the ask limit order will be matched with the bid order. Similarly, when sending a bid limit order, if there is an ask order on the order book that has the same or lower price than that of the bid order, then the bid limit order will be matched to the ask order. A market order is immediately matched with any existing order on the order book. When there are ask orders on the order book, any bid market order will be matched with the lowest priced ask order on the order book. Again, in the case when bid orders are on the order book, any incoming ask market order will be matched with the highest priced bid order on the order book. For matching the orders, the price priority rule is used. On the basis of the price priority rule, the priced highest bid order on the order book will be given priority over all other bid orders, and the lowest priced ask order on the order book will be given priority over all other ask orders. Another important rule is the time priority rule. If there are multiple orders on the order book with the same price, the oldest order is given priory against all other orders at that price, and will be executed first. Major exchanges around the world use both the price priority rule and the time priority rule. The trading price is the price of either the bid or ask order that was on the order book first.

We will next introduce several models that capture the continuous double auction mechanism. In this study, a stochastic model that uses the continuous double auction mechanism to capture the trading process is called the ``stochastic order-book model." One of the pioneers of the ``stochastic order-book model" is the model introduced by S. Maslov \cite{rev18}. In this early model, limit order and market order were chosen with equal probability. Bid and ask orders were chosen with equal probability as well. The limit order price was selected by a uniform random number within a certain range from the last price. This is a very simple model, but it successfully captures the power law in the cumulative frequency distribution of the price difference gathered through simulations. Since this model, various stochastic order-book models have been proposed. For example, in Maslov's model, the order price is selected by a uniform random number; however, other models have looked at certain circumstances and used a selection trend or some distribution for selection of the order price \cite{rev1}\cite{rev2}\cite{rev6}\cite{rev14}\cite{rev16}. These models can be broadly classified into two categories: one is the exclusion model where only one unit of order can be pooled in the same price, and another is the particle model where several orders can be pooled in the same price. In addition, Maskawa proposed a model where investors mimic the historical order trend \cite{rev11}. In this model, with a certain probability, the priority given to an order's price is based on the price that has the greatest pooled order volume on the order book. Simulations using this model demonstrate price shift, price gap, and power law in the spread. In addition, G. Harras and D. Sornette proposed a model that is not a stochastic order-book model, although the model captures the swarm behaviors of investors well \cite{rev7}. This model is based on the Ising model. Moreover, other studies have investigated agency-based models focusing on dealer behavior \cite{rev8}\cite{rev13}. In addition to simulations, numerous studies have conducted empirical analysis using actual market data \cite{rev15}\cite{rev20}.

In this study, we present a simple stochastic order-book model that incorporates the swarm behaviors of investors. Specifically, two swarm behaviors are incorporated in the model; one is investors' following a trend in the historical price movement, and another is investors' trading against a trend in the historical price movement. We denote the first model of investors following the historical trend as the ``following model." We denote the latter model, in which investors trade against the historical trend, as the ``contrary model." The model that does not include these swarm behaviors is denoted as the ``plain model."

We will next explain the basic structure of our model, that is, how the orders are selected. First, existing orders on the order book are examined to see whether there is an order at a price beyond a certain range from the base price. The base price is the latest execution price. If there is an order outside the range, the order will be removed from the order book. If there is no such order, then a new order will be placed on the order book. The new order will be either a bid or an ask order by $\frac{1}{2}$ probability and the price of the order will be randomly selected within a certain range from the base price. In this study, we empirically chose a range of $\pm 15$ from the base price. Therefore, if the base price is $0$, the bid or ask order price is randomly selected in the range $[-15, 15]$, and one unit will be placed on the order book. In this study, the signs of the numbers have no meaning. In these respects, our model is similar to Maslov's model; however, it should be noted that our model significantly differs in its order expiration mechanism \cite{rev18}. In Maslov's model, orders were removed from the order book if they were not executed within a certain time frame. However, in our model, if an order is outside the established range from the base price because the base price has moved away from the order price, then the order is removed from the order book. We use this mechanism because in reality, investing information is abundantly and readily available to investors; therefore, it is unlikely that their orders would be left on the order book when the base price has moved sufficiently away from their order price. In addition, in markets led by professional traders, traders are constantly calculating the theoretical price of a product; therefore, the entire trading community has similar ideas regarding appropriate pricing. Therefore, it is more realistic to remove an the order whose price is placed outside the established range from the base price. Furthermore, the stochastic order-book model, removing orders from the order book is significant. If orders other than executed orders are not removed from the order book, after a time, the balance between ask and bid orders on the order book can experience a distortion resulting in a price movement that can carry a strong one-sided trend. This movement differs from that of real markets. As is widely known, actual exchange markets generally regress to the means, and the order expiration mechanism is significant for enforcing regression to the means. Moreover, in this study, we employ the price priority rule that is used in the trading mechanism of the major global exchanges. The time priority rule is not meaningful in our simulation because we do not distinguish between agents who send orders. Trading takes place whenever best ask $\le$ best bid, where ``best ask" is the lowest ask price on the order book and "best bid" is the highest bid price on the order book. The transaction price is either the price of the bid or ask order, whichever is on the order book first. Figure $1$ is an example of a successful transaction. In our model, we do not use the market order. However, because an order is always placed in terms of one unit only, when an order is immediately executed, it could be interpreted as being a market order because it has having the same effect as a market order. The plain model follows the rules that have thus far been outlined.

We incorporated two types of investors' swarm behavior in the plain model. The ``following model" is for the follower behavior where investors trade in accordance with a trend in the historical price movement, and the ``contrary model" is for contrarian behavior where investors trade against a trend in the historical price movement. To model these behaviors, ``the price range" from the base price used in the ``plain model" is broken down into three ranges: first is price range $+6$ to $+15$ from the base price, second is price range $\pm5$ from base price and third is price range $-6$ to $-15$ from base price. Usually, an order will be randomly placed within the $\pm15$ price range as in the ``plain model." However, when the historical price movement has a certain trend, orders will be placed within the three ranges with different probabilities in accordance with the trend in the historical price movement. The past $10$ execution-price movements are used as the historical price movement. In the following model, if the price increases more than $7$ times within the past $10$ historical price movements, both bid and ask orders will more likely be placed within the $+6$ to $+15$ price range. Specifically, when the price is in the upward trend, the probability of placement will be $0.8$ in the $+6$ to $+15$ price range and $0.1$ in both the $\pm5$ and the $-6$ to $-15$ price ranges. This difference of in probability expresses how investors consider a trend in the historical price movement and ``swarm" to a certain price range. On the contrary, in the past $10$ price movements, if the price declines more than $7$ times, both bid and ask orders will be placed within the $-6$ to $-15$ price range with a $0.8$ probability, $0.1$ will be applied as the order probability in the other two ranges. Under Maskawa's model, orders were placed on a price it had the most orders on the order book. The orders were placed to this price with a certain probability \cite{rev11}. Our model captures how investors are affected by the historical price trend and accordingly swarm to price ranges higher or lower than the base price. Unlike Maskawa, we do not consider the order balance on the order book when an order is sent, but our model assumes that past price dynamics affect the investing tendency. Typically, investors make investment decisions based on historical analysis; therefore, our model incorporates actual practices. In the contrary model, investor behavior is the opposite of that assumed by the following model. In this case, when the past $10$ price movements include a price increase of more than $7$ times, both bid and ask orders will be placed within the $-6$ to $-15$ price range with a $0.8$ probability. When the past $10$ price movements include a price decline of more than $7$ times, then the order probability in the $+6$ to $+15$ price range will be $0.8$. The above mechanisms are used to capture investors swarming to a certain price range, modeling both when investors take a follower action and trade in accordance with the historical price movements, and when investors take contrarian actions, trading against the historical price movements. Past experimental studies have examined the distribution of order frequency in relation to price and showed that, the more the price moves away from the base price, the less frequent orders become, following a power law relationship \cite{rev9}\cite{rev10}\cite{rev12}. Whereas, in this research, for the purpose of defining the extent to which the swarm behaviors of investors will affect the price, the range of orders is simply divided into three categories that characterize the swarm behaviors.

\section{Parameters}
This section will describe the parameters of the simulations and their meaning in greater detail. The parameters required for the simulations are listed in Tables $1$ and $2$, where BP is the base price.

\begin{table}[h]
\begin{center}
\begin{tabular}{c|c|c|c}
 \hline
\ & \multicolumn{3}{c}{Possible order price range (BP = base price)} \\
\cline{2-4}
\ & (i) BP -15 $\sim$ BP -6  & (i) BP -5 $\sim$ BP +5  & (iii) BP +6 $\sim$ BP +15 \\
 \hline \hline
Following Model & 0.1 & 0.1 & 0.8 \\
 \hline
Contrary Model & 0.8 & 0.1 & 0.1\\
 \hline
\end{tabular}
\end{center}
\caption[Table] 
{The order probability in the next step for each price range when the past price increased more than $7$ out of $10$ times. The order price is randomly chosen within the selected price range.}
\label{table:Table1}
\end{table}

\begin{table}[h]
\begin{center}
\begin{tabular}{c|c|c|c}
 \hline
\ & \multicolumn{3}{c}{Possible order price range (BP = base price)} \\
\cline{2-4}
\ & (i) BP -15 $\sim$ BP -6  & (i) BP -5 $\sim$ BP +5  & (iii) BP +6 $\sim$ BP +15 \\
 \hline \hline
Following Model & 0.8 & 0.1 & 0.1 \\
 \hline
Contrary Model & 0.1 & 0.1 & 0.8\\
 \hline
\end{tabular}
\end{center}
\caption[Table] 
{The order probability in the next step for each price range when the past price declined more than $7$ out of $10$ times. The order price is chosen at random within the selected price range.}
\label{table:Table2}
\end{table}

In the plain model, orders are randomly placed within a range of $\pm 15$. In the following model and the contrary model, the parameters of the order probability for each price range are designed to generate and emphasize the swarm behaviors, and we can expect that, if the order probability for each price range was different, the effect on price would probably change. However, it is appropriate to attach more weight to price range conditions (i) and (iii) because the main purpose of this research is to explore the extent to which prices are affected by orders following the past price or by swarming in the opposite direction. The price range established between $\pm 15$ of the base price was empirically defined as such because, if the price range is wider, the price movement increases but there is no significant effect on swarm behavior, and if the price range is too narrow, it proves more difficult to characterize the swarm behaviors from the simulations. The tick count of the observed past trends was also empirically chosen with consideration for the frequency of the swarm behaviors appearing in each number of simulations. Both the tick count of the observed past trends and the price range where orders can be placed exhibit a neutral effect on the observed swarming behaviors.

\section{Simulation Results}
In this study, we analyzed the extent to which the price continuously moved in a single direction. We did not differentiate between downside or upside price movements. Instead, we used data regarding the size of the price movement when the price movement was in one direction continuously; that is, we captured the absolute value of the continuous price movements. Therefore, in other words, we obtained price data from our simulation and plotted a frequency distribution of the absolute value of the draw down and draw up in the price, where draw down is the decline of the price when the prices declined continuously, and draw up is the increase of the price when the prices increased continuously. In this study, the absolute values of the draw down and draw up are called the draw size. For example, a $-50$ draw down and a $+50$ draw up are not differentiated because the absolute value is used. Draw size is an appropriate indicator for this study because our purpose is to capture continuous price movement risk.

First, we conducted simulations for each of the three models: the plain model, following model and, the contrary model. One million simulations were conducted $30$ times for each model, and transactions occurred with the ratio $29.05 \pm 0.07 \%$ for the number of each simulations of the three models. Figure $2$ shows $10,000$ ticks of price data. A tick is a unit of price movement.

To discover how prices diffuse with time, Figure $3$ shows the relationship between the standard deviation of the price gap and the time scale in a double-logarithmic graph. The dotted line is proportional to the one-half power of the time scale. The relationship $\sigma (\tau)$ between the standard deviation of the price gap and the time scale is as follows, where $\tau$ is the price gap.

\begin{eqnarray}
\sigma (\tau) \propto \tau^{H}, \ \ \nonumber
H = 0.5, \ \ \tau \ge 1. \nonumber
\end{eqnarray}
The coefficient $H$ is called the Hurst exponent. The autocorrelation function of the price gap is shown by Figure $4$. When the time lag is greater than $1$, the autocorrelation is nearly $0$.

The draw size was taken from this price data. For all three models, the number of draw size data points was $17.74 \pm 0.15 \%$ of the number of simulations. The ratio of swarm behavior to the number of transactions was $0.92 \pm 0.16 \%$ for the following model and $0.19 \pm 0.03 \%$ for the contrary model. Moreover, we plotted the cumulative frequency distribution of the draw size to analyze the three models. Figure $5$ is the double-logarithmic graph of the cumulative frequency distributions of the draw sizes derived from the three models. This graph shows that the differences in the cumulative frequency distribution of the draw size between the models can be seen in the tail.

Next, we analyzed the cumulative frequency distribution of the draw size from the plain model. Figure $6$ represents the cumulative frequency distributions that compare the draw size from the plain model (Plain) and the draw size from shuffled price gap data from the plain model (Plain Shuffle). The solid line represents a linearization of the original data from the plain model. The slope of this straight line is $-0.04$. Moreover, the dotted line represents a linearization of the shuffled price gap data from the plain model. The slope of this straight line is $-0.06$. The correlation $R^2$ of the both linearizations is $0.99$. Figure $7$ is the semilogarithmic graph for the tail of the cumulative frequency distribution of the draw size from the plain model. The draw size data for the tail include all the data of every class that falls below $0.5\%$ of the largest draw size, even when only a part of the class is actually within the largest $0.5\%$. Here these data represent a draw size of $41$ and larger. The dotted line in the figure is the cumulative distribution function of a log normal distribution created by the maximum-likelihood method. We used the following equation to model the tail of the log normal distribution \cite{rev22}.

\begin{eqnarray}
P_{>}(x) &=& m \int^{\infty}_x C(\alpha, \beta) \frac{1}{x} \exp \biggl\{ -\alpha \log \Bigl( \frac{x}{s} \Bigl) - \beta \Bigl(\log \Bigl( \frac{x}{s} \Bigl) \Bigl)^2 \biggl\} dx, \nonumber \\
where \ \ \ \ C(\alpha, \beta) &=& \biggl[ \sqrt{\frac{\pi}{\beta}} e^{\frac{\alpha^2}{4\beta}} \Bigl( 1 - \Phi \Bigl(\frac{\alpha}{\sqrt{2\beta}} \Bigl) \Bigl) \biggl]^{-1}. \nonumber
\end{eqnarray}
Here $P_{>}(x)$ is the probability of certain data being $x$ or above, $m$ is the ratio of the number of data $x \ (x \geq s)$ against the number of all data, and $s$ is the smallest draw size of the tail. In addition, the function $\Phi$ is the cumulative distribution function of a standard normal distribution. For the log normal distribution of this graph, the best fit parameters are $\alpha = 3.89$ and $\beta = 1.77$.

Next, we analyzed the cumulative frequency distribution of the draw size from the following model. Figure $8$ represents the cumulative frequency distributions that compare the draw size from the following model (Following) and the draw size from shuffled price gap data from the following model (Following Shuffle). The chain line represents a linearization of the shuffled price gap data from the following model. The slope of this straight line is $-0.06$. The correlation $R^2$ of the linearization is $0.99$. Moreover, as Figure $9$ shows, we compared the plain model and the following model using data for which draw size is $16$ or larger. As you can see from the graph, the results are similar up till where draw size is close to $40$. On the other hand, where the draw size is larger than $40$, large draw size can be observed more frequently in the following model than in the plain model.

In order to see the influence of swarm behaviors, we then analyzed the shape of the tail of cumulative frequency distribution of the draw size from the following model. Figure $10$ is the double-logarithmic graph for the tail of the cumulative frequency distribution of the draw size from the following model. As in the plain model, this graph includes all draw size data of all classes in the largest $0.5\%$. The dotted line is the cumulative distribution function of a log normal distribution created so that the sum of squares of the difference from the data would be minimum. Here we use this method because we could not get an appropriate parameter by the maximum-likelihood method. For the log normal distribution of this graph, the best fit parameters are $\alpha = 4.00$ and $\beta = 0.13$. The solid line is the cumulative distribution function of a power law distribution created by the maximum-likelihood method. This cumulative distribution function follows the formula below.

\begin{eqnarray}
P_{>}(x) \propto x^{-4.19}. \nonumber
\end{eqnarray}
The maximum-likelihood estimate of the power index of the simulations conducted in this study was $4.19 \pm 0.31$.

Finally, we will analyze the extent to which the distributions of the draw size from each model differ. We compared the differences between each swarm behavior's effect on price. We analyzed the deviations of the cumulative frequency distributions of the draw size of the following model and the contrary model from the cumulative frequency distribution of the draw size of the plain model. Specifically, we used draw size data of $16$ and larger, calculated the difference of each cumulative frequency distribution, and compared the sum of squares. The sum of square difference between the cumulative frequency distributions of the following model and the plain model was more than $13$ times larger than that between the cumulative frequency distributions of the contrary model and the plain model. In addition, the Kolmogorov-Smirnov test was applied to determine the difference, if any, between these distributions. We perform the test using a class range of $15$. The cumulative relative frequency of the draw size from the three models are listed in Table $3$. Also, the statistics of Kolmogorov-Smirnov test are listed in Table $4$.

\begin{table}[hbtp]
 \label{table:data_type}
 \begin{center}
  \begin{tabular}{r|ccc}
   \hline
    Class & Plain & Following & Contrary \\
   \hline \hline
   150 & $1.87 \times 10^{-7}$ & $2.18 \times 10^{-5}$ & - \\
   135 & $7.50 \times 10^{-7}$ & $3.71 \times 10^{-5}$ & $1.87 \times 10^{-7}$ \\
   120 & $2.25 \times 10^{-6}$ & $7.30 \times 10^{-5}$ & $7.50 \times 10^{-7}$ \\
   105 & $9.37 \times 10^{-6}$ & $1.36 \times 10^{-4}$ & $4.31 \times 10^{-6}$ \\
   90 & $4.33 \times 10^{-5}$ & $2.51 \times 10^{-4}$ & $2.72 \times 10^{-5}$ \\
   75 & $1.94 \times 10^{-4}$ & $5.15 \times 10^{-4}$ & $1.43 \times 10^{-4}$ \\
   60 & $8.33 \times 10^{-4}$ & $1.28 \times 10^{-3}$ & $7.54 \times 10^{-4}$ \\
   45 & $3.64 \times 10^{-3}$ & $4.13 \times 10^{-3}$ & $3.57 \times 10^{-3}$ \\
   30 & $1.56 \times 10^{-2}$ & $1.63 \times 10^{-2}$ & $1.58 \times 10^{-2}$ \\
   15 & $7.83 \times 10^{-2}$ & $7.90 \times 10^{-2}$ & $7.87 \times 10^{-2}$ \\
   0 & $1.00$ & $1.00$ & $1.00$ \\
   \hline
  \end{tabular}
 \end{center}
 \caption{Cumulative relative frequency of the draw size from the three models}
\end{table}

\begin{table}[hbtp]
 \label{table:data_type}
 \begin{center}
  \begin{tabular}{l|c}
   \hline
     & Statistics \\
   \hline \hline
   Plain and Following & $4.79$ \\
   Following and Contrary & $3.33$ \\
   Contrary and Plain & $1.25$ \\
   \hline
  \end{tabular}
 \end{center}
 \caption{Statistics of Kolmogorov-Smirnov test}
\end{table}

\section{Discussion of the Numerical Results}
In this section, we examine the results of the empirical analysis in the previous section. First, we compared the price movements of the three models. The price movements shown in Figure $2$ indicate that, compared to that in the plain model or contrary model, the price movement in the following model is a large-scale continuous movement. We can assume that this phenomenon is induced by investors' swarming behavior to follow the historical trend. It also suggests that our model adequately reflects the swarm behavior incorporated in the plain model. On the other hand, for the contrary model we found little difference from the behavior of the plain model, even in the price movement graph.

Figure $3$ shows that the standard deviation of the price gap for the time scale is proportional to the one-half power of the time scale for each model. This is similar to the results obtained from past experimental studies  \cite{rev1} \cite{rev11}, which reported that slow diffusion occurred over a short time scale, where the Hurst exponent was below $0.5$, and approached a value of $0.5$ at a longer time scale. The data derived from the current simulations, however, exhibited an absence of slow diffusion over short time scales. This indicates that the price data from each model has the same diffusion speed representative of a random walk. Moreover, in the autocorrelation functions of the price gaps for the three models shown in Figure $4$, the time series data show no long-term memory. 

Next, we studied the common features of the draw size cumulative frequency distribution of the three models. Figure $5$ shows that the distribution shapes substantially is deviate from around a draw size of $16$. We think this is because a draw size of $16$ and larger reflects only the effects of pure consecutive price movements in a single direction.

Next, we looked at the plain model. Figure $3$, Figure $6$ and Figure $7$ suggest that the price data of the plain model resemble a random walk. If a random walk is assumed in a market, a given draw size will occurs exponentially less frequently when its size is larger  \cite{rev4}. Moreover, Figure $6$ shows that, the slope of the linearized data is different, but the cumulative frequency distribution of the draw size from the plain model share features with the cumulative frequency distribution of the draw size from the shuffled price gap data from the plain model. Therefore, the  plain model corresponds to an ideal market that resembles a random walk. Thus, by comparing the plain model with models incorporating swarm behavior, we can analyze the effect of swarm behavior.

We then examined the following model. Figure $8$ shows the diversion of Following and Following Shuffle. This indicates that the price movements from the following model have the emergent temporal correlations. Moreover, Figure $5$ and Figure $9$ show that, the larger the draw size, the wider the tail of the cumulative frequency distribution. Assuming that the price data from the plain model are like a random walk, we think that price data from the following model exhibit autocorrelation. Our model successfully captures the effect of swarm behavior following a historical trend.

Additionally, in order to catch the shape of the tail of the cumulative frequency distribution  of the draw size from the following model, we compared the cumulative frequency distribution from the model with a log normal distribution, and we also compared the cumulative frequency distribution from the model with a power law distribution. As Figure $10$ shows, when we estimate the shape of the distribution with consideration to the data at around the edge of the tail, the tail of the cumulative distribution obtained from the draw size of the following model fits better to a log normal distribution than power law distribution. However, aside from the data at around the edge of the tail, there was no significant difference between a log normal distribution and power law distribution. Generally, a log normal distribution and a power law distribution are confused easily. The probability density function of a log normal distribution can be expressed by the following formula,

\begin{eqnarray}
p(x) &=& \frac{1}{\sqrt{2 \pi \sigma^2}} \frac{1}{x} \exp \Bigl \{ -\frac{(\log x - \mu)^2}{2 \sigma^2}\Bigl \}, \nonumber \\
&=& \frac{1}{\sqrt{2 \pi \sigma^2}} \frac{1}{e^{\mu}} \Bigl( \frac{x}{e^{\mu}} \Bigl)^{-1-\xi(x)}, \nonumber \\
where \ \ \ \xi(x) &=& \frac{1}{2\sigma^2} \log \Bigl( \frac{x}{e^{\mu}} \Bigl). \nonumber
\end{eqnarray}
The reason for confusion is because $\xi(x)$ slowly moves against $x$ \cite{rev3}. Also, by looking at the edge of the data which is draw size $291$, we know that the estimated log normal distribution is not completely capturing the edge of the tail. In the model of this study, one tick can move maximum of $15$. Draw size $291$ shows that this model has inherent price movement risk that is about $20$ times larger than the maximum movement of one tick. This risk needs to be appropriately measured in order to manage the significant price decline risk. Therefore, estimation of distribution of around the edge of data has to be done cautiously. In anyway, we confirmed that the distribution is more fat tail than normal distribution of the data from the following model estimated. These show that investors'  swarm behaviors produced by investors behaviors interacting create a phenomenon similar to a significant price decline. Also, similar results to the simulation results presented here have been shown by previous research on the distribution of the return of the US stock market \cite{rev15}\cite{rev20}. From these past studies, it is known that a power law emerges from the distribution of the return of the US stock market. In particular, these studies show that the exponent is approximately $3$. In another study, it was shown that the distribution of the draw down and draw up obtained from actual market data has a fatter tail than the index distribution of the expected distribution that assumes a complete random walk market \cite{rev4}. Although the shape of the distribution is subject to further discussion, the results of previous empirical studies are similar to our simulation results; therefore, we suggest that multiple investors' behaviors interact and create a phenomenon that has a fat tail in the actual market as well. Thus, we think that in the real market, when there are many investors that trade by following the historical price movement, there will be events that the typically assumed random walk cannot capture.

Finally, we will consider how the distributions of the draw size differ from each model. As Figure $5$ shows, the cumulative frequency distribution of the draw size from the contrary model has a slightly tighter tail than that of the draw size from the plain model but no significant differences are observed. Moreover, calculating the differences between the cumulative frequency distributions of the draw size from the plain model and the following model, and the plain model and the contrary model and comparing the sum of square differences, the sum of square differences of the plain model and the following model is larger plainly. Intuitively, orders placed against the current trend would have a stabilization effect on price movement. However, these results show that swarming behaviors of order placements against the current trend do not have a significant influence on the frequency of the draw size. Therefore, the swarm behavior of investors following the price movement trend has a greater impact on the price than that of investors opposing the price movement trend. In addition, at a significance level of $10\%$, the result of the Kolmogorov-Smirnov test rejects the hypothesis that the distribution of the draw size derived from the plain model is identical to that from the following model. This indicates that there is a certain degree of difference between the distributions. However, can we not strongly conclude that the distributions of the draw size differ between the two models, having tested at a significance level of $10\%$. The test covered the entire distribution range and, therefore, the impact of small draw sizes that occur more frequently is reflected strongly in the result.

\section{Conclusion}
We successfully incorporated the swarm behaviors of investors in a simple stochastic order-book model. We expressed swarm behavior by adding a rule that changes the order inclination depending on the historical trend to the plain model that assumes investors to randomly send orders to the order book. In particular, a fat tail could be observed from the distribution of the draw size obtained from the following model incorporating the swarm behavior of investors that follow the trend of the past. This shows that, compared to a market that assumes price movement as a random walk, larger continuous price movements occur more frequently. In other words, numerous investors' behaviors interact and structure themselves as a swarm behavior and create a phenomenon similar to significant price decline. In fact, the result demonstrated that one of the reasons the trend following of price occurs is that the orders temporarily swarm on the order book in accordance with past price trends. On the other hand, the fat tail phenomenon has been observed in actual financial markets, such as power law in tail distribution of return. Thus, we found that the proposed, simple stochastic order-book model incorporating the swarm behaviors of investors, despite its simplicity, captures a real-world phenomenon. Also, from all the above, we found that, when investors swarm and send orders following a historical price movement, large-scale continuous price movements in a single direction, such as price declines, can occur.

In addition, when estimating the shape of the tail of distribution of draw size from the following model, we found that a log normal distribution fits better than the power law distribution, However, although log normal distribution is more applicable to the shape of the tail of distribution, it did not completely capture the edge of the distribution that shows the probability of phenomenon such as significant price decline. When managing risk with the consideration of the phenomenon, power law distribution may be more appropriate. In any case, when calculating return or risk in the financial market, careful discussion is required to appropriately assess the shape the tail of the distribution.

For future study, we could analyze how to handle the data of rare case that is at around the edge of the distribution, the method for estimating the distribution and conduct mathematical observation for capturing the phenomenon.  Moreover, a recent study discussed that the order book has a layered structure in the foreign exchange market, and that the structure is related to trend following and so on \cite{rev23}. We want to reproduce these advanced phenomena by improving our stochastic order-book model. These analyses are expected to effective in quantifying the risk of price movement.

\clearpage

\begin{figure}[!htb]
\begin{center}
\hspace{-0.4cm}
\includegraphics[clip,width=140mm]{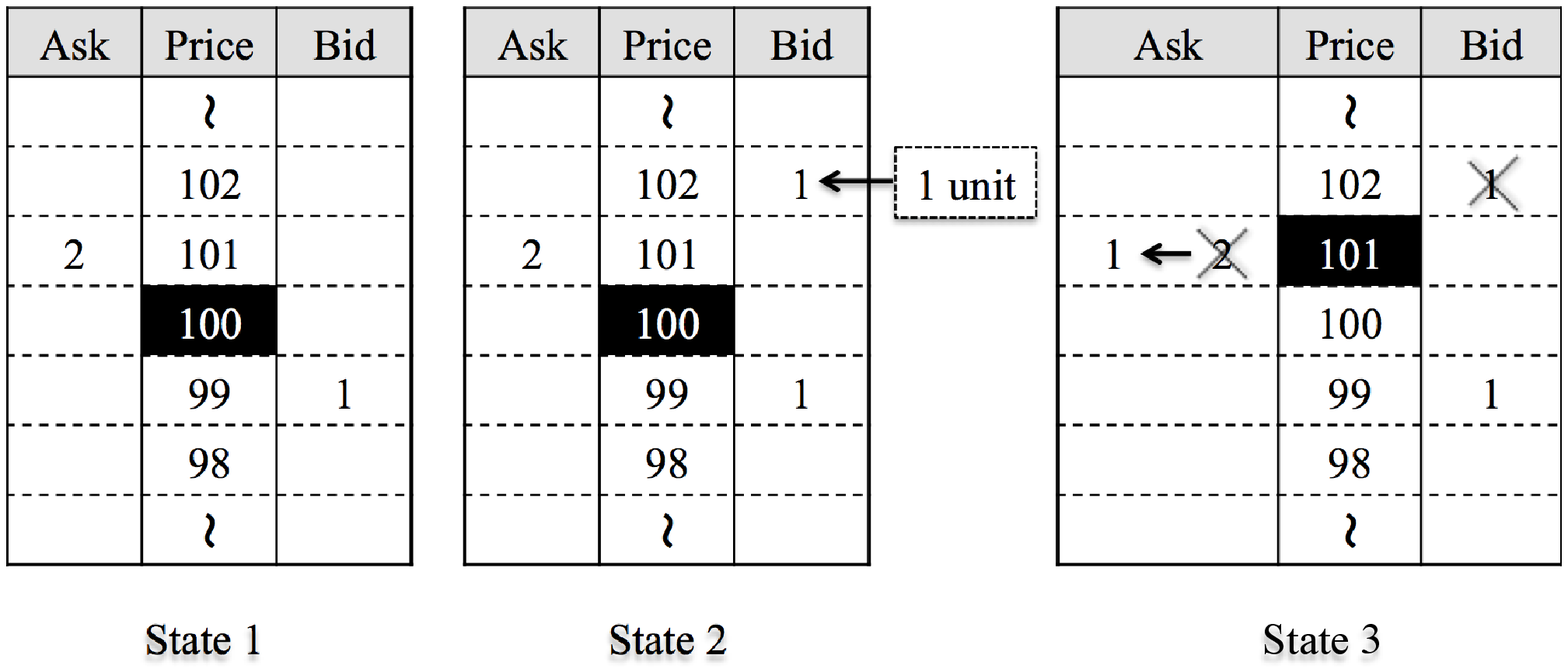}
\hspace{-0.8cm}
\end{center}
\caption[]{Drawing provides an example of an order-book transaction. At the starting point, the order book has one unit of sell order (bid) at a price of 99 and two units of buy order (ask) at a price of 101, and the base price is 100 (State $1$). Then, one unit of bid order at a price of 102 is entered (State $2$). Because of this new order, best ask $\le$ best bid; therefore, the transaction occurs between the one unit ask order at a price of 101 and the new bid order at a price of 102 (State $3$). The transaction price is the price of the order that was on the order book first, a price of 101. The Base Price changes to 101, because of the transaction.}
\label{figure:Fig1}
\end{figure}

\begin{figure}[!htb]
\begin{center}
\hspace{-0.4cm}
\includegraphics[clip,width=140mm]{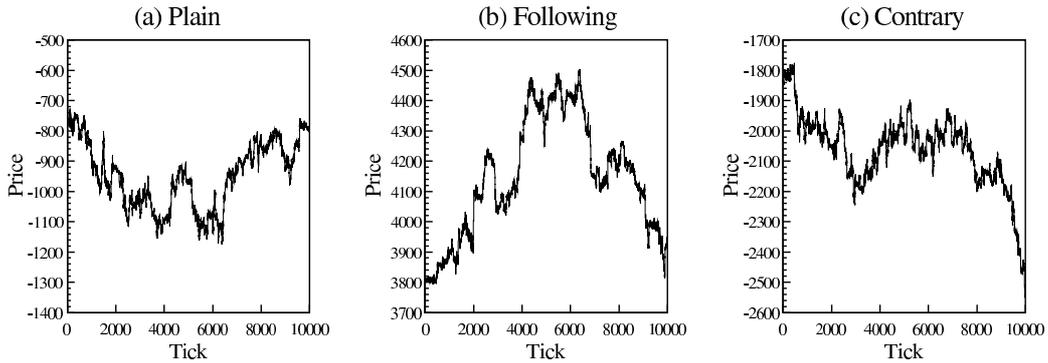}
\hspace{-0.8cm}
\end{center}
\caption[]{Price movement for 10,000 ticks. The price fluctuations are obtained by simulations using (a) the plain model, (b) the following model, and (c) the contrary model.}
\label{figure:Fig2}
\end{figure}

\begin{figure}[!htb]
\begin{center}
\hspace{-0.4cm}
\includegraphics[clip,width=100mm]{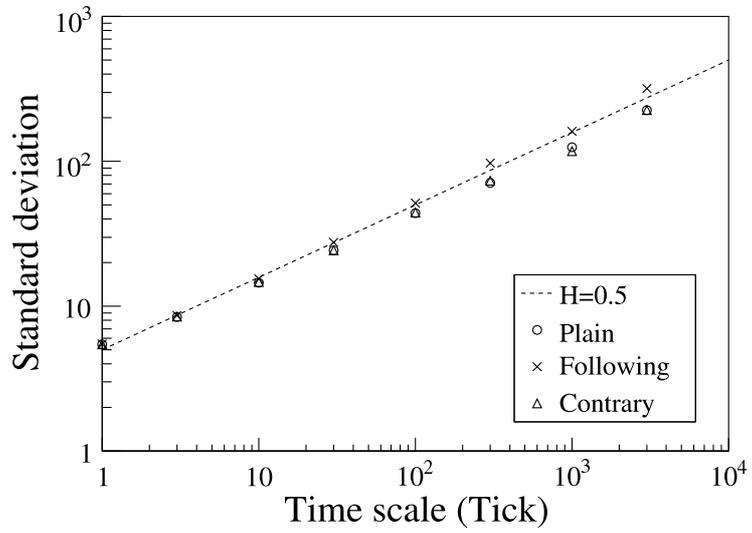}
\hspace{-0.8cm}
\end{center}
\caption[]{Double-logarithmic graph of the standard deviations of the price gaps with respect to the time scale derived from the three models. The dotted line is the Hurst exponent of $0.5$.}
\label{figure:Fig3}
\end{figure}

\begin{figure}[!htb]
\begin{center}
\hspace{-0.4cm}
\includegraphics[clip,width=100mm]{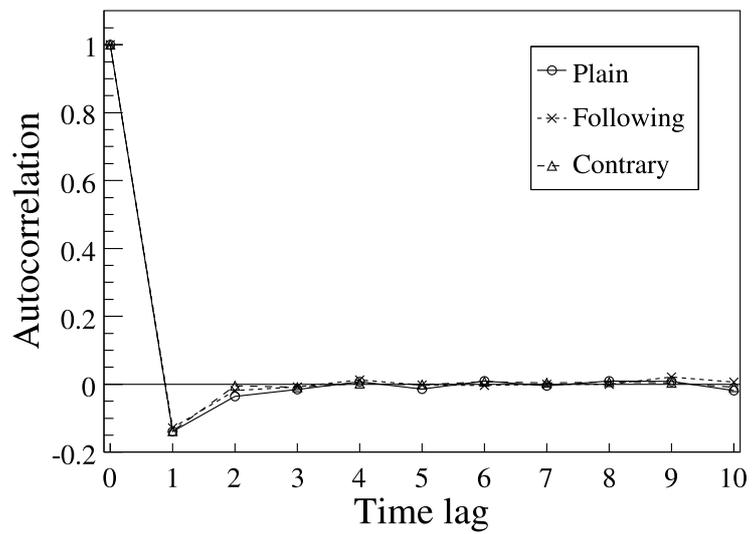}
\hspace{-0.8cm}
\end{center}
\caption[]{Autocorrelation functions of the price gap derived from the three models.}
\label{figure:Fig4}
\end{figure}

\begin{figure}[!htb]
\begin{center}
\hspace{-0.4cm}
\includegraphics[clip,width=100mm]{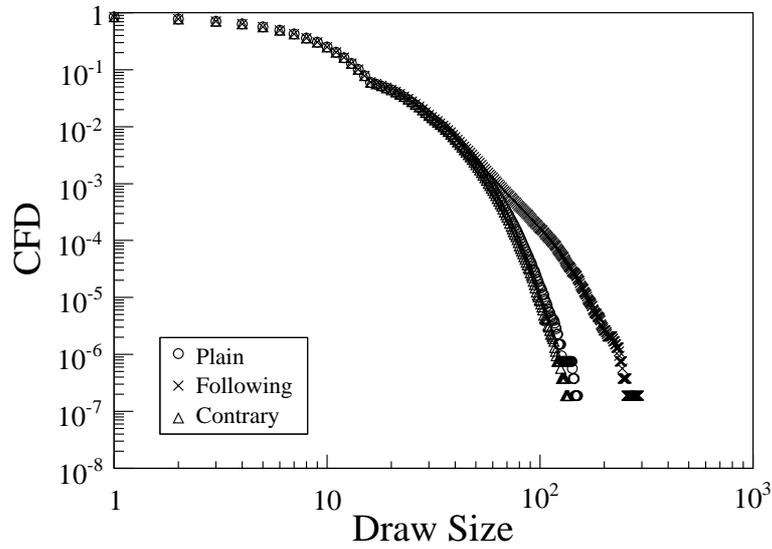}
\hspace{-0.8cm}
\end{center}
\caption[]{Cumulative frequency distributions of the draw size from the three models.}
\label{figure:Fig5}
\end{figure}

\begin{figure}[!htb]
\begin{center}
\hspace{-0.4cm}
\includegraphics[clip,width=100mm]{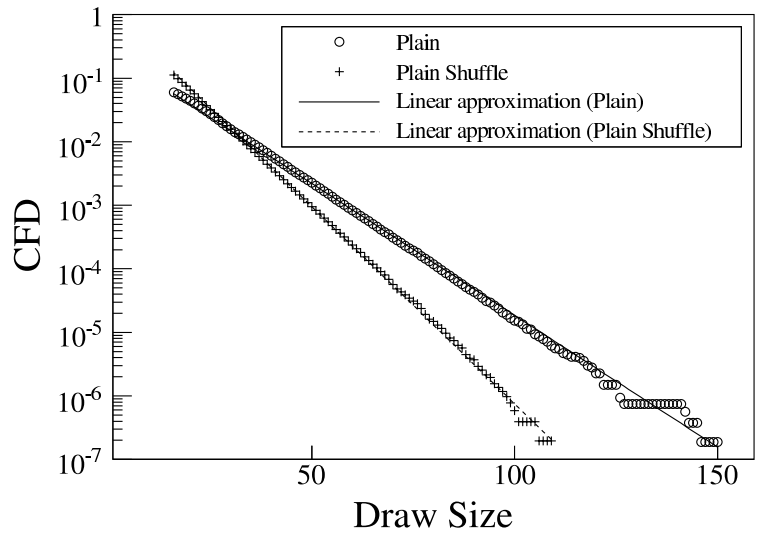}
\hspace{-0.8cm}
\end{center}
\caption[]{Semilogarithmic graph of the cumulative frequency distributions for a draw size of $16$ and larger from the plain model (Plain) and from the shuffled price gap data from the plain model (Plain Shuffle). The solid line represents a linearization of the original data from the plain model and has a slope of $-0.04$. The dotted line represents a linearization of the shuffled price gap data from the plain model and has a slope of $-0.06$. The correlation $R^2$ of the both linearizations is $0.99$.}
\label{figure:Fig6}
\end{figure}

\begin{figure}[!htb]
\begin{center}
\hspace{-0.4cm}
\includegraphics[clip,width=100mm]{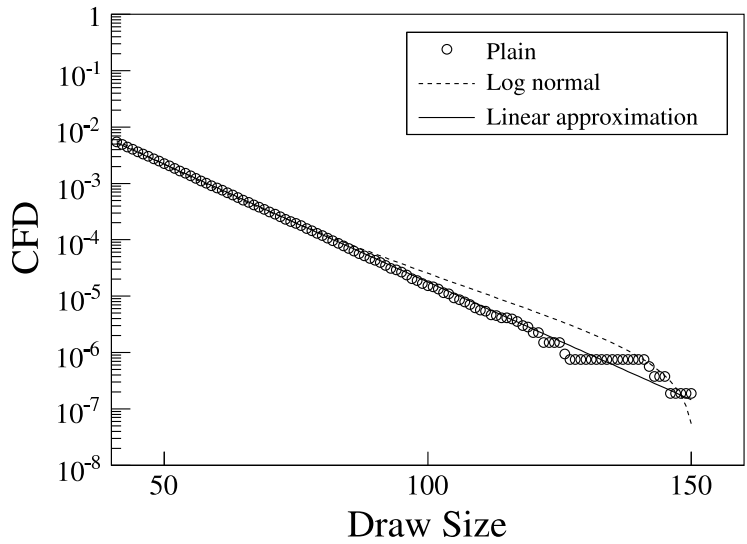}
\hspace{-0.8cm}
\end{center}
\caption[]{Semilogarithmic graph for the tail of the cumulative frequency distribution of the draw size from the plain model. The dotted line is the best fit using the cumulative of a log normal distribution. The best fit parameters are $\alpha = 3.89$ and $\beta = 1.77$. The solid line is the same as the solid line of Figure $6$.}
\label{figure:Fig7}
\end{figure}

\begin{figure}[!htb]
\begin{center}
\hspace{-0.4cm}
\includegraphics[clip,width=100mm]{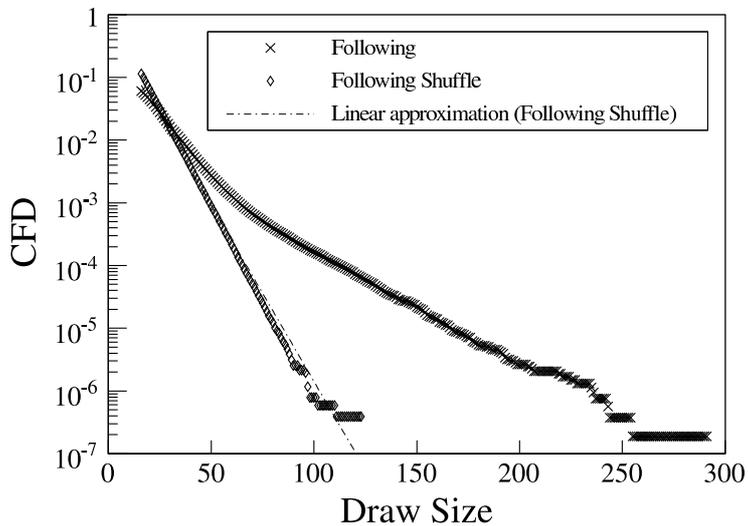}
\hspace{-0.8cm}
\end{center}
\caption[]{Semilogarithmic graph of the cumulative frequency distributions for a draw size of $16$ and larger from the following model (Following) and from the shuffled price gap data from the following model (Following Shuffle). The chain line represents a linearization of the shuffled price gap data from the following model and has a slope of $-0.06$. The correlation $R^2$ of the linearization is $0.99$.}
\label{figure:Fig8}
\end{figure}

\begin{figure}[!htb]
\begin{center}
\hspace{-0.4cm}
\includegraphics[clip,width=100mm]{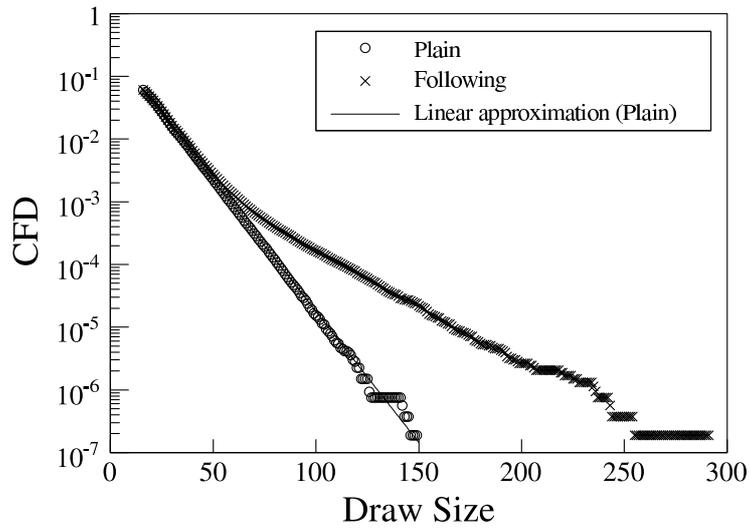}
\hspace{-0.8cm}
\end{center}
\caption[]{Semilogarithmic graph for the tail of the cumulative frequency distribution of a draw size of $16$ and larger from the plain model and the following model. The solid line is the same as the solid line of Figure $6$.}
\label{figure:Fig9}
\end{figure}

\begin{figure}[!htb]
\begin{center}
\hspace{-0.4cm}
\includegraphics[clip,width=100mm]{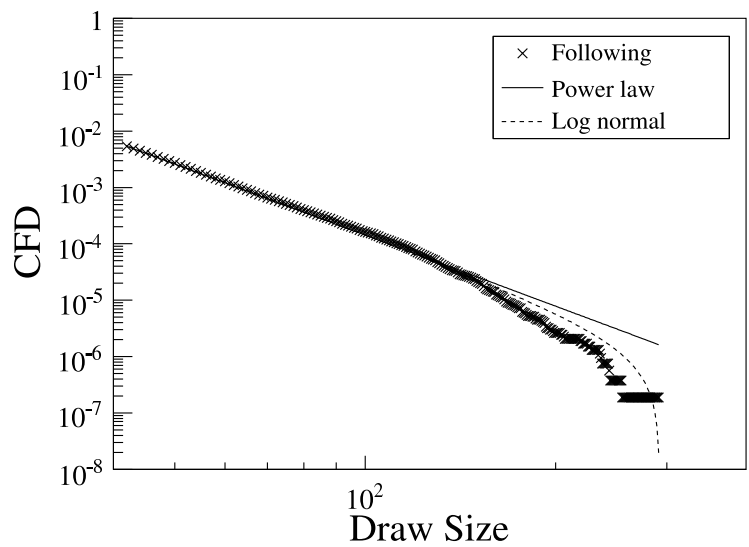}
\hspace{-0.8cm}
\end{center}
\caption[]{Double-logarithmic graph for the tail of the cumulative frequency distribution of the draw size from the following model. The dotted line is the best fit using the cumulative of a log normal distribution. The best fit parameters are $\alpha = 4.00$ and $\beta = 0.13$. The solid line is the best fit using the cumulative distribution of a power law. This maximum-likelihood estimate of the power index is $4.19$.}
\label{figure:Fig10}
\end{figure}

\end{document}